\documentclass[showpacs,superscriptaddress]{revtex4}
\usepackage{amssymb,amsmath}
\usepackage[dvips,final]{graphicx}
\usepackage{url}
\usepackage{subfigure}
\usepackage{textcomp}
\usepackage{graphicx}
\usepackage{bm}
\usepackage{dcolumn}
\usepackage[usenames]{color}
\usepackage{epsfig}
\input epsf.tex
\usepackage{amssymb}
\usepackage{amsmath}
\usepackage{bm}
\usepackage{amsthm}
\usepackage{braket}
\usepackage[title]{appendix}
\usepackage{amsopn}
\usepackage{hyperref}

\def\comment#1{}

\def\beq{\begin{equation}}
\def\eeq{\end{equation}}
\def\bea{\begin{eqnarray}}
\def\eea{\end{eqnarray}}

\makeatletter
\usepackage{ulem}
\usepackage{cancel}
\usepackage{color}

\makeatother

\begin{document}

\title{Photon-graviton scattering: A new way to detect anisotropic gravitational waves?}

\author{Nicola Bartolo}
\email[]{ nicola.bartolo@pd.infn.it}

\affiliation{Dipartimento di Fisica e Astronomia ``Galileo Galilei", Universit\`a degli Studi di Padova, via Marzolo
8, I-35131, Padova, Italy}
\affiliation{INFN Sezione di Padova, via Marzolo 8, I-35131, Padova, Italy}
\affiliation{INAF-Osservatorio Astronomico di Padova, vicolo dell'Osservatorio 5, I-35122 Padova, Italy}

\author{Ahmad Hoseinpour}
\email[]{ahmad.hoseinpour@ph.iut.ac.ir}

\affiliation{Department of Physics, Isfahan University of Technology, Isfahan
84156-83111, Iran}
\affiliation{ICRANet-Isfahan, Isfahan University of Technology, Isfahan 84156-83111, Iran}

\author{Giorgio Orlando}
\email[]{  giorgio.orlando@phd.unipd.it}

\affiliation{Dipartimento di Fisica e Astronomia ``Galileo Galilei", Universit\`a degli Studi di Padova, via Marzolo
8, I-35131, Padova, Italy}
\affiliation{INFN Sezione di Padova, via Marzolo 8, I-35131, Padova, Italy}

\author{Sabino Matarrese}
\email[]{sabino.matarrese@pd.infn.it }

\affiliation{Dipartimento di Fisica e Astronomia ``Galileo Galilei", Universit\`a degli Studi di Padova, via Marzolo
8, I-35131, Padova, Italy}
\affiliation{INFN Sezione di Padova, via Marzolo 8, I-35131, Padova, Italy}
\affiliation{INAF-Osservatorio Astronomico di Padova, vicolo dell'Osservatorio 5, I-35122 Padova, Italy}

\author{Moslem Zarei}
\email[]{m.zarei@cc.iut.ac.ir}

\affiliation{Department of Physics, Isfahan University of Technology, Isfahan
84156-83111, Iran}
\affiliation{ICRANet-Isfahan, Isfahan University of Technology, Isfahan 84156-83111, Iran}

\date{\today}

\date{\today}

 \begin{abstract}

 \noindent Gravitons are the quantum counterparts of gravitational waves in low-energy theories of gravity. Using Feynman rules one can compute scattering amplitudes describing the interaction between gravitons and other fields. Here, we consider the interaction between gravitons and photons. Using the quantum Boltzmann equation formalism, we derive fully general equations describing the radiation transfer of photon polarization, due to the forward scattering with gravitons. We show that the Q and U  photon linear polarization modes couple with the V photon circular polarization mode, if gravitons have anisotropies in their power-spectrum statistics. As an example, we apply our results to the case of primordial gravitons, considering models of inflation where an anisotropic primordial graviton distribution is produced. Finally, we evaluate the effect on cosmic microwave background (CMB) polarization, showing that in general the expected effects on the observable CMB frequencies are very small. However, our result is promising, since it could provide a novel tool for detecting anisotropic backgrounds of gravitational waves, as well as for getting further insight on the physics of gravitational waves.

 \end{abstract}


\maketitle

\section{Introduction}

\noindent From an effective field theory (EFT) point of view, general relativity can be considered as the low energy limit of a well-defined renormalizable theory of gravity. In this low energy theory, gravitational waves are (transverse-traceless) fluctuations around a suitable background space-time; expanding the Einstein-Hilbert Lagrangian in a power series of these fluctuations we are able to extract a positive-definite kinetic term, describing the free propagation of gravitational waves. Thus, we are able to canonically quantize gravitational waves, introducing creation and annihilation operators, as it is done with the particle content and gauge bosons of the standard model of fundamental interactions. In such a formalism field operators describing gravitational waves are called gravitons. In particular, expanding the full covariant Lagrangian of a generic theory in power series of all the field content, we can derive interaction terms between gravitons and other fields.

Here, using this quantum field theory (QFT) approach, we study the radiation transfer of a photon due to the so-called ``gravitational Compton scattering" with gravitons. In this case the full Lagrangian we have to consider is the sum of the Einstein-Hilbert Lagrangian and the covariant electromagnetic Lagrangian. Feynman amplitudes describing gravitational Compton scattering have already been computed in the  literature (see e.g. Refs. \cite{Skobelev1975, Bohr:2015, Bohr:2017}). The time evolution of the polarization matrix of a photon is given by the formalism of the so-called ``quantum Boltzmann equation" (see e.g. Refs. \cite{Kosowsky:1994cy,Alexander:2009,Zarei:2010,Mohammadi:2014, Zarei:2015, Sadegh:2017}). It is possible to show that this formalism is equivalent at lowest order in scattering kinematics to the classical radiative transfer, hence it provides a more general framework. In particular, here we will focus on the so-called ``forward scattering" term of the quantum Boltzmann equation. In fact, this term appears to be the lowest-order term in a perturbative expansion in power series of the interaction Hamiltonian of the theory, thus it is expected to provide the most sensitive effects on the dynamics of the photon polarization.

In this paper we will show that when a photon interacts with gravitons that are characterized by anisotropies in their power-spectrum statistics, the photon's Stokes parameters $Q$ and $U$ couple with the $V$ Stokes parameter in a non-trivial way, causing a change in the state of polarization of the photon. In particular, we will apply our result to the interaction between anisotropic primordial gravitons from inflation and cosmic microwave background (CMB) photons. We will show that, differently from the classical description (see e.g. Refs. \cite{Hu:1997,dodelson:2003}), this effect may lead to the production of the $V$ polarization state for CMB photons. However, in our case it turns out that in general the amplitude of such a polarization is expected to be very small. This makes the effect very difficult to detect through CMB measurements. The possibility of having $V$ mode generation in the CMB is an issue already studied in the literature, see e.g. Refs. \cite{Alexander:2009, Zarei:2010, Mohammadi:2014, King:2016,Ejlli:2016, Ejlli:2017, Sadegh:2017, Spider:2017, Montero:2018}; in particular, in Ref. \cite{King:2016} detection prospects are discussed, and in Ref. \cite{Spider:2017} some observational constraints are derived.

Despite the smallness of the effect we studied, we believe that the main result of our analysis is promising: on the one hand because one can try to look for this effect, by considering different frameworks for the photon-graviton (forward-scattering) interactions, and, on the other hand, because one could develop controlled photon beams, to constrain the presence of 
anisotropic gravitational-wave backgrounds in the Universe.

The paper is organized as follows. In Sec. II we study the effect of the photon-graviton forward scattering in the photon polarization, deriving fully general Boltzmann equations for the photon. In Sec. III we employ this set of equations in the context of inflationary models where an anisotropic background of gravitational waves is produced; we also perform a general estimate of the detectability of the expected effects on the polarization of CMB photons. Finally, in Sec. IV  we summarize our work and briefly comment on the importance of the current result for future applications.

\section{Effects of photon-graviton forward scattering on photon polarization}

\subsection{Forward scattering of photon with graviton}

\noindent The quantum Boltzmann equation for a photon interacting with a graviton is given by \cite{Kosowsky:1994cy}

\bea
(2\pi)^3 \delta^{(3)}(0)(2k^0)
\frac{d\rho^{(\gamma)}_{ij}(k)}{dt} = i\left\langle \left[H_{\gamma g}
(t),\mathcal{D}^{(\gamma)}_{ij}(k)\right]\right\rangle - \frac{1}{2} \int_{- \infty}^{+\infty} dt' \left\langle  \left[H_{\gamma g}
(t'), \left[H_{\gamma g}
(t),\mathcal{D}^{(\gamma)}_{ij}(k)\right] \right]\right\rangle ~\;,\label{Boltzmann_equation1} 
\eea
where $\rho^{(\gamma)}_{ij}$ is the polarization matrix of the photon [see Eq. \eqref{gamma_pol}], $\mathcal{D}^{(\gamma)}_{ij} = a_i^\dagger a_j$  is the photon
number operator and $H_{\gamma g}(t)$ is the \textit{quantum} interaction Hamiltonian between photons and gravitons, defined in terms of the second order S-matrix as \cite{Kosowsky:1994cy}
\beq
S^{(2)} =  -i\int_{-\infty}^{\infty} dt \, H_{\gamma g}(t)~\;.\label{intH} 
\eeq
The first term on the right-hand side of Eq. \eqref{Boltzmann_equation1} is the so-called forward scattering term, while the second term is the so-called damping or nonforward scattering term. Equation \eqref{Boltzmann_equation1} is obtained adopting a perturbative approach so that increasing powers of the interaction Hamiltonian $H_{\gamma g}(t)$ reduce the strength of the corresponding term. For this reason, we drop the damping term and we consider only the forward scattering one. Thus, our Boltzmann equation reads
\bea
(2\pi)^3 \delta^{(3)}(0)(2k^0)
\frac{d\rho^{(\gamma)}_{ij}(k)}{dt} = i\left\langle \left[H_{\gamma g}
(t),\mathcal{D}^{(\gamma)}_{ij}(k)\right]\right\rangle~.\label{Boltzmann equation2}
\eea
In order to derive the photon-graviton interaction Hamiltonian we follow the same convention of Refs. \cite{Skobelev1973,Skobelev1975}.
The gravitational dynamics is given by the Einstein-Hilbert
Lagrangian 
\beq
\mathcal{L}_{g}=\frac{\sqrt{-g}}{\kappa^2}g^{\alpha\lambda}\left(\Gamma^{\beta}_{\alpha\lambda}\Gamma^{\mu}_{\beta\mu}-\Gamma^{\mu}_{\alpha\beta}
\Gamma^{\beta}_{\lambda\mu}\right)~,\label{Lg}
\eeq
where $\kappa^2=16\pi G$ and we have dropped, as usual, irrelevant surface terms. We consider the weak-field limit and expand the metric around Minkowski space-time in powers of $\kappa$, as
\beq \label{weak_field}
g_{\mu\nu}(x)=\eta_{\mu\nu}+\kappa \, h_{\mu\nu}(x) \;,
\eeq
where $h_{\mu\nu}(x)$ is the graviton field. Now, we have to expand the Lagrangian \eqref{Lg} around the background in powers of $\kappa$ and $h_{\mu\nu}(x)$. The terms linear in $h_{\mu\nu}(x)$ vanish by virtue of Einstein's equations. We keep only the nonlinear terms in $h_{\mu\nu}(x)$, up to first order in $\kappa$. In fact, it is possible to show that terms with higher powers of 
$\kappa$ give a gradually suppressed contribution, when inserted into Eq. \eqref{Boltzmann equation2} . Thus, we have
\beq
\mathcal{L}_{g}=\mathcal{L}^{(0)}_{g}+\mathcal{L}^{(1)}_{g}+\mathcal{O}(\kappa^2) \;,
\eeq
where $\mathcal{L}^{(0)}_{g}$ is the Lagrangian describing the free propagation of gravitons and reads
\beq
\mathcal{L}^{(0)}_{g}=\frac{1}{4}\left\{h^{\alpha,\beta}_{\alpha}\left(2h^{,\lambda}_{\beta\lambda}-h^{\alpha}_{\alpha,\beta}\right)+
h^{\sigma\lambda,\alpha}\left(-2h_{\alpha\lambda,\sigma}+h_{\sigma\lambda,\alpha}\right)
\right\} \, ,\label{Lg0}
\eeq
$\mathcal{L}^{(1)}_{g}$ is the three-gravitons interaction Lagrangian and is given by
\bea
\mathcal{L}^{(1)}_{g}&=&\kappa\left\{\frac{1}{2}h^{\alpha}_{\alpha}\mathcal{L}^{(0)}_{g}
-\frac{1}{4}h^{\lambda\rho}\left[2h^{\alpha}_{\alpha,\sigma}\left(h^{\sigma}_{\lambda,\rho}-h_{\lambda \rho}^{,\sigma}\right)+
\left(-2h_{\alpha \lambda,\sigma} h^{\sigma,\alpha}_{\rho}+2h_{\sigma \lambda,\alpha} h^{\sigma,\alpha}_{\rho}
+h_{\alpha\sigma,\lambda}h^{ \sigma \alpha}_{,\rho}\right)
\right.\right. \nonumber \\ && \left.\left.
+h^{\alpha}_{\alpha,\lambda}\left(2h^{\nu}_{\rho,\nu}-h^{\alpha}_{\alpha,\rho}\right)
+2h^{\sigma\nu}_{,\nu}h_{\lambda\rho,\sigma}-4h^{\sigma,\alpha}_{\rho} h_{\alpha \sigma,\lambda}
\right]
\right\}.
\eea
Here and in the following Greek indices are raised and lowered using the Minkowski metric $\eta_{\mu\nu}$. Now, we make the following spatial Fourier expansion for the graviton field
\beq\label{quant}
h_{\mu\nu}(x)=\int\frac{d^{3}q}{(2\pi)^{3}}\frac{1}{2q^{0}}\sum_{r=+,\times}\left[b_r(\mathbf{q})\,h^{(r)}_{\mu\nu}\,e^{iqx}+
b^{\dag}_{r'}(\mathbf{q})\,h^{(r)\,\ast}_{\mu\nu}\,e^{-iqx}\right] \;,
\eeq
where $b^{(r)}_{\mathbf{q}}$ and $b^{(r)\,\dag}_{\mathbf{q}}$ are graviton annihilation and creation operators, respectively, obeying
the canonical commutation relations
\beq \label{quant_2}
\left[b_r(\mathbf{q}),b^{\dag}_{r'}(\mathbf{q}')\right]=(2\pi)^32q^0\delta^{(3)}(\mathbf{q}-\mathbf{q}')\delta_{rr'}~,
\eeq
and $h^{(r)}_{\mu\nu}$ are the polarization tensors with the well-known properties
\beq
h^{(r)}_{\mu\nu}(q)q^{\mu}=0~,\:\:\:\:\:\:\:\:h_{\mu}^{\mu}(q)=0~,\:\:\:\:\:\:\:\:h^{(r)}_{\mu\nu}(q)\left(h^{(r')\,\mu\nu}(q)\right)^{\ast}= \delta^{r r'}~ \label{canonical}.
\eeq
It is also convenient to represent the polarization tensor $h^{(r)\,\mu\nu}$ in terms of a direct product of unit spin polarization vectors
\beq
h^{(r)}_{\mu\nu}=e^{(r)}_{\mu}e^{(r)}_{\nu}~,\:\:\:\:\:\:\:e^{(r)}_{\mu}q^{\mu}=0 ~,\:\:\:\:\:\:\:
  \left[ e^{(r)}_{\mu}(q) \left(e^{(r')\,\mu}(q)\right)^{\ast} \right]^2=  \delta^{r r'}~.
\eeq
The explicit expression for the graviton propagator in the harmonic (de Donder) gauge is given by \cite{Skobelev1973,Skobelev1975}
\beq
D^{(g)}_{\mu\nu\alpha\beta}(q)=\frac{i}{q^2}\left(\eta_{\mu\alpha}\eta_{\nu\beta}+
\eta_{\mu\beta}\eta_{\nu\alpha}-\eta_{\mu\nu}\eta_{\alpha\beta}
\right)~.
\eeq
To compute the S-matrix \eqref{intH} we need to consider also the covariant electromagnetic Lagrangian density \cite{Skobelev1973,Skobelev1975}
\beq
\mathcal{L}_{\gamma}=\mathcal{L}^{(0)}_{\gamma}+\mathcal{L}^{(1)}_{\gamma}+\mathcal{L}^{(2)}_{\gamma}+\mathcal{O}(\kappa^3)~, \label{photonL}
\eeq
where $\mathcal{L}^{(0)}_{\gamma}$ is the part of the Lagrangian that describes the free photon propagation and is given by
\beq
\mathcal{L}^{(0)}_{\gamma}=-\frac{1}{16\pi}F_{\mu\nu}F^{\mu\nu} \;,
\eeq
where, as usual, $F_{\mu\nu}=\partial_{\mu}A_{\nu}-\partial_{\nu}A_{\mu}$ is the gauge-field strength and $A^{\mu}$ is the photon field. The remaining terms on the right-hand side of Eq. \eqref{photonL} give the photon-graviton couplings
\beq
\mathcal{L}^{(1)}_{\gamma}=-\frac{\kappa}{16\pi}\left(\frac{1}{2}h_{\alpha}^{\alpha}
\,F_{\mu\nu}F^{\mu\nu}-2h^{\mu\nu}F_{\mu}^{\alpha}F_{\alpha\nu}\right)~,
\eeq
and
\bea \label{Lg2}
\mathcal{L}^{(2)}_{\gamma}=-\frac{\kappa^{2}}{16\pi}\left\{-h_{\alpha}^{\alpha}h^{\mu\nu}F_{\mu}^{\alpha}F_{\alpha\nu}
+\left[\frac{1}{8}\left(h_{\alpha}^{\alpha}\right)^{2}-\frac{1}{4}h_{\mu\nu}h^{\mu\nu}\right]F_{\alpha\beta}F^{\alpha\beta}
+h^{\mu\nu}h^{\alpha\beta}F_{\mu\alpha}F_{\nu\beta}+
2h^{\mu\alpha}h^{\nu}_{\alpha}F_{\mu\beta}F_{\nu}^{\beta}\right\}~.
\eea
In particular, $\mathcal{L}^{(1)}_{\gamma}$ gives the two photons-one graviton interaction vertex, while $\mathcal{L}^{(2)}_{\gamma}$ gives the two photons-two gravitons vertex. Notice that in Eq. \eqref{photonL} we considered also a coupling term which is quadratic in $\kappa$. This should not create any confusion, since the important fact is that, when we will apply the formula \eqref{Boltzmann equation2}, the term \eqref{Lg2} will give a contribution of the same order in powers of $\kappa$. 

As we made for the graviton field, we expand also the photon field in Fourier space as
\beq
A_{\mu}(x)=\int\frac{d^{3}p}{(2\pi)^{3}}\frac{1}{2p^{0}}\sum_{s=1,2}\left[a_s(\mathbf{p})\,\epsilon^{(s)}_{\mu}\,e^{ipx}+
a^{\dag}_{s'}(\mathbf{p})\,\epsilon^{(s)\,\ast}_{\mu }\,e^{-ipx}\right] \;,
\eeq
where $\epsilon^{(s)}_{\mu}$ are the real photon polarization
four-vectors, $s$ labels the two physical transverse photon polarizations, $a_{s}$ and $a^{\dag}_{s'}$ are photon annihilation and creation operators, respectively, satisfying
the canonical commutation relation
\beq
\left[a_s(\mathbf{p}),a^{\dag}_{s'}(\mathbf{p}')\right]=(2\pi)^32p^0\delta^{(3)}(\mathbf{p}-\mathbf{p}')\delta_{ss'} \;.
\eeq
We fix the Feynman gauge for the photon field, thus the photon propagator is given by \cite{Skobelev1973,Skobelev1975}
\beq
D^{(\gamma)}_{\mu\nu}(p)=-4\pi\frac{i}{p^2}\,\eta_{\mu\nu}~.
\eeq
Now, we have all the elements to evaluate the right-hand side of Eq. \eqref{Boltzmann equation2}. In fact, the expression of the second order S-matrix contribution describing the photon-graviton scattering is
\beq \label{S2}
S^{(2)}=-\frac{1}{2}\int d^{4}x_1d^{4}x_2 \, T\left\{  \mathcal{L}^{(1)}_{\gamma}(x_1)
\mathcal{L}^{(1)}_{\gamma}(x_2) \right\}-\frac{1}{2}\int d^{4}x_1 d^4x_2 \, T\left\{\mathcal{L}^{(1)}_{\gamma}(x_1)
\mathcal{L}^{(1)}_{g}(x_2)\right\}+i\int d^{4}x\,  \mathcal{L}^{(2)}_{\gamma}(x) \;,
\eeq
where T denotes the time-ordering operator.

\begin{figure}
  \includegraphics[width=4.5in]{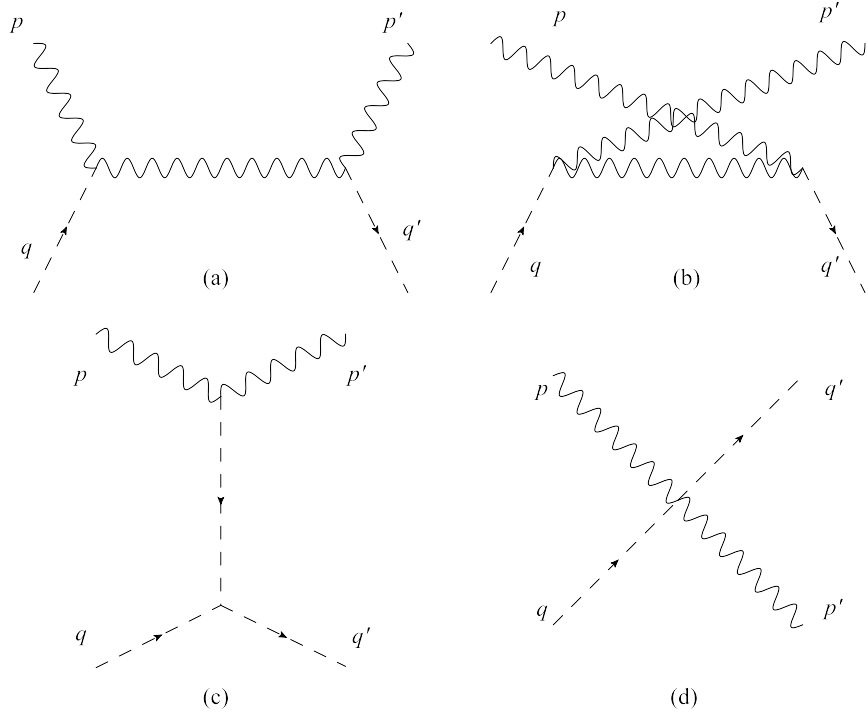}
\caption{Feynman diagrams for the photon-graviton scattering, dashed lines represent gravitons, wavy lines represent photons. Diagrams (a) and (b) give the amplitude $M_1$, Eq. \eqref{m1}; 
diagram (c) gives the amplitude $M_2$, Eq. \eqref{m2}; diagram (d) gives the amplitude $M_3$, Eq. \eqref{m3}.}\label{feyndiag}
\end{figure}

Now, calling $p$ and $p'$ the incoming and outgoing momenta of the photon, $q$ and $q'$ the incoming and outgoing momenta of the graviton, we can evaluate the second-order S-matrix \eqref{S2} using Feynman's rules (see Refs. \cite{Skobelev1975, Bohr:2015}). In particular, Feynman diagrams for the photon-graviton scattering are shown in Fig. \ref{feyndiag}. The result is such that the photon-graviton interaction Hamiltonian defined in \eqref{intH} turns out to be
\bea \label{H}
H_{\gamma g}(t)&=&\int d\mathbf{q}d\mathbf{q}'d\mathbf{p}d\mathbf{p}'(2\pi)^{3}\delta^{(3)}(\mathbf{q}'+\mathbf{p}'-\mathbf{q}-\mathbf{p})\exp\left[it(q'^0+p'^0-q^0-p^0)\right] \nonumber \\&& \times \left[b^{\dag}_{r'}(\mathbf{q}')a^{\dag}_{s'}(\mathbf{p}')(M_1+M_2+M_3)a_{s}(\mathbf{p})b_{r}(\mathbf{q})\right] \;,
\eea
where for simplicity of notation
 \beq
d\mathbf{q}=\frac{d^{3}\mathbf{q}}{(2\pi)^{3}2q^{0}}~, \:\:\:\:\;\:\:\:\:\:\:\:\:\:\;\:\:\:\:\:
d\mathbf{p}=\frac{d^{3}\mathbf{p}}{(2\pi)^{3}2p^{0}}~,
\eeq
 and the three different Feynman amplitudes appearing in Eq. \eqref{H} are given by \cite{Skobelev1975}

 \bea \label{m1}
 M_1&=&\frac{\kappa^2}{p\cdot q}\left[p^{\mu}(e\cdot\epsilon)-\epsilon^{\mu}(p\cdot e)\right]\left[p'^{\nu}(\epsilon'^{\ast}\cdot e'^{\ast})
 -\epsilon'^{\ast\nu}(p'\cdot e'^{\ast})\right]
 \nonumber \\&& \times
 \left[g^{(0)}_{\mu\nu}(e\cdot p)(e'^{\ast}\cdot p')+q_{\mu}q'_{\nu}(e\cdot e'^{\ast})
 -e'^{\ast}_{\mu}q'_{\nu}(e\cdot p)-q_{\mu}e_{\nu}(e'^{\ast}\cdot p')
 \right]+(\epsilon,p\leftrightarrow \epsilon'^{\ast},-p')~,
 \eea

\bea \label{m2}
 M_2&=&\frac{\kappa^2}{2q\cdot q'} \left\{\left[(p\cdot p')(\epsilon\cdot\epsilon'^{\ast})-(\epsilon'^{\ast}\cdot p)
 (\epsilon\cdot p')\right][2(e\cdot q')(e'^{\ast}q)-(e\cdot e'^{\ast})(q\cdot q')](e\cdot e'^{\ast})
 \right. \nonumber \\ && \left.
 -\left[(p\cdot p')\epsilon^{\mu}\epsilon'^{\ast\nu}+(\epsilon\cdot \epsilon'^{\ast})p^{\mu}p'^{\nu}-(p\cdot \epsilon'^{\ast})\epsilon^{\mu}p'^{\nu}
 -(\epsilon\cdot p')p^{\mu}\epsilon'^{\ast\nu}+(\epsilon,p\leftrightarrow \epsilon'^{\ast},-p')\right]
 \right. \nonumber \\ && \left.
 \times
 \left[(e\cdot e'^{\ast})^{2}q_{\mu}q'_{\nu}-2(e\cdot e'^{\ast})(q\cdot q')e_{\mu}e'^{\ast}_{\nu}
 +2(e\cdot q')(e'^{\ast}\cdot q)e_{\mu}e'^{\ast}_{\nu}+(e\cdot q')^{2}e'^{\ast}_{\mu}e'^{\ast}_{\nu}
  \right. \right. \nonumber \\ && \left. \left.
 +
 (e'^{\ast}\cdot q)^2 e_{\mu}e_{\nu}-2(e\cdot e'^{\ast})(e\cdot q')e'^{\ast}_{\nu}q'_{\mu}
 -2(e\cdot e'^{\ast})(e'^{\ast}\cdot q )e_{\nu}q_{\mu} \right]
     \right\}~,
 \eea

\bea \label{m3}
 M_3&=& \kappa^2 \left \{ (e\cdot e'^{\ast})^{2}[(p\cdot p')(\epsilon\cdot \epsilon'^{\ast})-
 (p\cdot \epsilon'^{\ast})(p'\cdot \epsilon)]-2[(p\cdot e)(\epsilon \cdot e'^{\ast})
 -(p\cdot e'^{\ast})(\epsilon\cdot e)]
 \right. \nonumber \\ && \left.
 \times[(p'\cdot e)(\epsilon'^{\ast}\cdot e'^{\ast})-(p'\cdot e'^{\ast})(\epsilon'^{\ast}\cdot e)]
    -2(e\cdot e'^{\ast})[(p\cdot e)(p'\cdot e'^{\ast})(\epsilon\cdot \epsilon'^{\ast})
 \right. \nonumber \\ && \left. +(\epsilon\cdot e)(\epsilon'^{\ast}\cdot e'^{\ast})(p\cdot p')
  -(p\cdot e)(\epsilon'^{\ast}\cdot e'^{\ast})(\epsilon\cdot p')
  -(e\cdot \epsilon)(p'\cdot e'^{\ast})(p\cdot \epsilon'^{\ast})
   \right. \nonumber \\ && \left.
  +(\epsilon,p\leftrightarrow \epsilon'^{\ast},-p')
  ]
    \right\} \;,
 \eea
where $e\equiv e^{(r)}(q)$, $e'\equiv e^{(r')}(q')$, $\epsilon\equiv \epsilon^{(s)}(p)$  and $\epsilon'\equiv \epsilon^{(s')}(p')$. At the end one can explicitly show that the contribution of $M_1+M_3$ amplitudes in the forward scattering vanishes. Hence, we focus only on $M_2$. After expanding $M_2$ and doing some algebra we obtain
\bea
   M_2&=&-\frac{\kappa^2}{2(q\cdot q')}\left\{(e\cdot e'^{\ast})^2\left[(q\cdot \epsilon'^{\ast})\left((p\cdot p')(\epsilon \cdot q')-(p \cdot q')(\epsilon \cdot p')\right)+(\epsilon \cdot \epsilon'^{\ast}) \left((p\cdot q')(q\cdot p')+(p\cdot p')(q\cdot q')\right. \right.\right.  \nonumber \\ && \left. \left.\left.
+(p\cdot q)(p'\cdot q') \right) -(p\cdot \epsilon'^{\ast})\left((q\cdot q')(\epsilon\cdot p')+(q\cdot p')(\epsilon\cdot q')+(q\cdot \epsilon)(p'\cdot q')\right)+(q'\cdot \epsilon'^{\ast})\left((p\cdot p')(q\cdot \epsilon)\right.\right. \right.\nonumber \\ && \left. \left.\left.
-(p\cdot q)(\epsilon \cdot p')\right)\right]- 2 (e\cdot e'^{\ast})\left\{-(e\cdot p')(p\cdot \epsilon'^{\ast})(q\cdot \epsilon)(q\cdot e'^{\ast})-(e\cdot \epsilon)(p\cdot \epsilon'^{\ast})(q\cdot p')(q\cdot e'^{\ast})\right. \right.  \nonumber \\ && \left. \left.
+(e\cdot \epsilon)(p\cdot p')(q\cdot \epsilon'^{\ast})(q\cdot e'^{\ast})-(e\cdot q')(p\cdot \epsilon'^{\ast})(\epsilon \cdot p')(q\cdot e'^{\ast})-(e\cdot p)(q\cdot \epsilon'^{\ast})(\epsilon\cdot p')(q\cdot e'^{\ast})\right.  \right. \nonumber \\ && \left. \left.
-(e\cdot p')(p\cdot \epsilon'^{\ast})(q\cdot q')(\epsilon\cdot e'^{\ast})+(e\cdot \epsilon'^{\ast})\left[(p\cdot p')\left((q\cdot \epsilon)(q\cdot e'^{\ast})+(q\cdot q')(\epsilon \cdot e'^{\ast})\right)-(\epsilon\cdot p')\left((p\cdot q)(q\cdot e'^{\ast}) \right. \right.\right.\right.  \nonumber \\ && \left.\left. \left.\left.
+(p\cdot e'^{\ast})(q\cdot q')\right)\right]-(e\cdot \epsilon)(p\cdot \epsilon'^{\ast})(q\cdot q')(p'\cdot e'^{\ast})-(e\cdot q')(p\cdot \epsilon'^{\ast})(\epsilon\cdot q')(p'\cdot e'^{\ast})+(e\cdot \epsilon)(p\cdot p')(q\cdot q')(e'^{\ast}\cdot \epsilon'^{\ast})\right.\right.   \nonumber \\ &&\left. \left.
-(e\cdot q')(p\cdot q')(\epsilon\cdot p')(e'^{\ast}\cdot \epsilon'^{\ast})-(e\cdot p)(q\cdot q')(\epsilon\cdot p')(e'^{\ast}\cdot \epsilon'^{\ast})+(e\cdot q')(p\cdot p')(\epsilon\cdot q')(e'^{\ast}\cdot \epsilon'^{\ast})\right. \right.  \nonumber \\ && \left.\left.
-(e\cdot q')(p\cdot \epsilon'^{\ast})(\epsilon\cdot e'^{\ast})(p'\cdot q')+(\epsilon \cdot \epsilon'^{\ast})\left[(e\cdot p')\left((p\cdot q)(q\cdot e'^{\ast})+(p\cdot e'^{\ast})(q\cdot q')\right)+(e\cdot p)\left((q\cdot e'^{\ast})(q\cdot p')\right. \right. \right. \right.  \nonumber \\ && \left.\left.\left. \left.
+(q\cdot q')(e'^{\ast}\cdot p')\right)+(e\cdot q')\left((p\cdot p')(q\cdot e'^{\ast})+(p\cdot q')(e'^{\ast}\cdot p')+(p\cdot e'^{\ast})(p'\cdot q')\right)\right]+(e\cdot q')(p\cdot p')(\epsilon\cdot e'^{\ast})(q'\cdot \epsilon'^{\ast}) \right.\right. \nonumber\\&& \left.\left.
-(e\cdot q')(p\cdot e'^{\ast})(\epsilon\cdot p')(q'\cdot \epsilon'^{\ast})\right\}+2\left\{(e\cdot q')^2\left[(p'\cdot e'^{\ast})\left((\epsilon \cdot \epsilon'^{\ast}) (p\cdot e'^{\ast})-(p\cdot \epsilon'^{\ast})(\epsilon\cdot e'^{\ast})\right)\right. \right.\right. \nonumber\\&& \left. \left. \left.
+(e'^{\ast}\cdot \epsilon'^{\ast})\left((p\cdot p')(\epsilon\cdot e'^{\ast})-(p\cdot e'^{\ast})(\epsilon\cdot p')\right)\right]+(q\cdot e'^{\ast})(e\cdot q')\left[(e\cdot p')\left((\epsilon \cdot \epsilon'^{\ast})(p\cdot e'^{\ast})-(p\cdot \epsilon'^{\ast})(\epsilon\cdot e'^{\ast})\right)\right. \right. \right. \nonumber\\&& \left. \left. \left.
+(e\cdot \epsilon'^{\ast})\left((p\cdot p')(\epsilon\cdot e'^{\ast})-(p\cdot e'^{\ast})(\epsilon\cdot p')\right)+(e\cdot p)\left((\epsilon \cdot \epsilon'^{\ast})(e'^{\ast}\cdot p')-(\epsilon\cdot p')(e'^{\ast}\cdot \epsilon'^{\ast})\right)\right]\right. \right.  \nonumber\\&& \left. \left.
+(e\cdot p)(q\cdot e'^{\ast})^2\left((\epsilon \cdot \epsilon'^{\ast})(e\cdot p')-(e\cdot \epsilon'^{\ast})(\epsilon\cdot p')\right)+(e\cdot \epsilon)(q\cdot e'^{\ast})\left[(e\cdot \epsilon'^{\ast})(p\cdot p')(q\cdot e'^{\ast})\right.\right.\right.\nonumber\\&&\left.\left.\left.
-(p\cdot \epsilon'^{\ast})\left((e\cdot p')(q\cdot e'^{\ast})+(e\cdot q')(e'^{\ast}\cdot p')\right)+(e\cdot q')(p\cdot p')(e'^{\ast}\cdot \epsilon'^{\ast})\right]\right\}\right\}~.
\eea
Now, using the four-momentum conservation $q\cdot q'=p\cdot p'$, we can rewrite $M_2$ as
\bea
   M_2&=&-\frac{\kappa^2}{2}\left\{(e\cdot e'^{\ast})^2\left((q\cdot \epsilon'^{\ast})(\epsilon \cdot q')+(\epsilon \cdot \epsilon'^{\ast})(p\cdot p')-(p\cdot \epsilon'^{\ast})(\epsilon\cdot p')+(q'\cdot \epsilon'^{\ast})(q\cdot \epsilon)\right)- 2 (e\cdot e'^{\ast})\left[(e\cdot \epsilon)(q\cdot \epsilon'^{\ast})(q\cdot e'^{\ast}) \right.\right. \nonumber\\&&\left.\left.
-(e\cdot p')(p\cdot \epsilon'^{\ast})(\epsilon\cdot e'^{\ast})+(e\cdot \epsilon'^{\ast})\left((q\cdot \epsilon)(q\cdot e'^{\ast})+(p\cdot p')(\epsilon \cdot e'^{\ast})-(\epsilon\cdot p')(p\cdot e'^{\ast})\right)-(e\cdot \epsilon)(p\cdot \epsilon'^{\ast})(p'\cdot e'^{\ast})\right.\right. \nonumber\\&&\left.\left.
+(e\cdot \epsilon)(p\cdot p')(e'^{\ast}\cdot \epsilon'^{\ast})-(e\cdot p)(\epsilon\cdot p')(e'^{\ast}\cdot \epsilon'^{\ast})+(e\cdot q')(\epsilon\cdot q')(e'^{\ast}\cdot \epsilon'^{\ast})+(\epsilon \cdot \epsilon'^{\ast})\left((e\cdot p')(p\cdot e'^{\ast})\right.\right.\right. \nonumber\\&&\left.\left.\left.
+(e\cdot p)(e'^{\ast}\cdot p')+(e\cdot q')(q\cdot e'^{\ast})\right)+(e\cdot q')(\epsilon\cdot e'^{\ast})(q'\cdot \epsilon'^{\ast})\right]+2\left[(e\cdot q')^2(e'^{\ast}\cdot \epsilon'^{\ast})(\epsilon\cdot e'^{\ast})\right.\right. \nonumber\\&&\left.\left.
+(q\cdot e'^{\ast})(e\cdot q')(e\cdot \epsilon'^{\ast})(\epsilon\cdot e'^{\ast})+(e\cdot \epsilon)(q\cdot e'^{\ast})\left((e\cdot \epsilon'^{\ast})(q\cdot e'^{\ast})+(e\cdot q')(e'^{\ast}\cdot \epsilon'^{\ast})\right)\right]\right\} \nonumber\\&&
-\frac{\kappa^2}{2(q\cdot q')}\left\{(e\cdot e'^{\ast})^2\left[-(q\cdot \epsilon'^{\ast})(p \cdot q')(\epsilon \cdot p')+(\epsilon \cdot \epsilon'^{\ast}) \left((p\cdot q')(q\cdot p')+(p\cdot q)(p'\cdot q') \right) -(p\cdot \epsilon'^{\ast})\left((q\cdot p')(\epsilon\cdot q')\right. \right.\right.  \nonumber \\ && \left. \left.\left.
+(q\cdot \epsilon)(p'\cdot q')\right)-(q'\cdot \epsilon'^{\ast})(p\cdot q)(\epsilon \cdot p')\right]- 2 (e\cdot e'^{\ast})\left\{-(e\cdot p')(p\cdot \epsilon'^{\ast})(q\cdot \epsilon)(q\cdot e'^{\ast})-(e\cdot \epsilon)(p\cdot \epsilon'^{\ast})(q\cdot p')(q\cdot e'^{\ast})\right.\right. \nonumber \\ && \left. \left.
-(e\cdot q')(p\cdot \epsilon'^{\ast})(\epsilon \cdot p')(q\cdot e'^{\ast})-(e\cdot p)(q\cdot \epsilon'^{\ast})(\epsilon\cdot p')(q\cdot e'^{\ast}) -(e\cdot \epsilon'^{\ast})(\epsilon\cdot p')(p\cdot q)(q\cdot e'^{\ast})\right. \right. \nonumber \\ && \left. \left.
-(e\cdot q')(p\cdot \epsilon'^{\ast})(\epsilon\cdot q')(p'\cdot e'^{\ast})-(e\cdot q')(p\cdot q')(\epsilon\cdot p')(e'^{\ast}\cdot \epsilon'^{\ast})-(e\cdot q')(p\cdot \epsilon'^{\ast})(\epsilon\cdot e'^{\ast})(p'\cdot q')\right. \right. \nonumber \\ && \left.\left.
+(\epsilon \cdot \epsilon'^{\ast})\left[(e\cdot p')(p\cdot q)(q\cdot e'^{\ast})+(e\cdot p)(q\cdot e'^{\ast})(q\cdot p')+(e\cdot q')\left((p\cdot q')(e'^{\ast}\cdot p')+(p\cdot e'^{\ast})(p'\cdot q')\right)\right] \right.\right. \nonumber\\&& \left.\left.
-(e\cdot q')(p\cdot e'^{\ast})(\epsilon\cdot p')(q'\cdot \epsilon'^{\ast})\right\}+2\left\{(e\cdot q')^2\left[(p'\cdot e'^{\ast})\left((\epsilon \cdot \epsilon'^{\ast}) (p\cdot e'^{\ast})-(p\cdot \epsilon'^{\ast})(\epsilon\cdot e'^{\ast})\right)\right. \right. \right. \nonumber\\&& \left. \left. \left.
-(e'^{\ast}\cdot \epsilon'^{\ast})(p\cdot e'^{\ast})(\epsilon\cdot p')\right]+(q\cdot e'^{\ast})(e\cdot q')\left[(e\cdot p')\left((\epsilon \cdot \epsilon'^{\ast})(p\cdot e'^{\ast})-(p\cdot \epsilon'^{\ast})(\epsilon\cdot e'^{\ast})\right)\right.\right.\right.\nonumber\\&&\left.\left.\left.
-(e\cdot \epsilon'^{\ast})(p\cdot e'^{\ast})(\epsilon\cdot p')+(e\cdot p)\left((\epsilon \cdot \epsilon'^{\ast})(e'^{\ast}\cdot p')-(\epsilon\cdot p')(e'^{\ast}\cdot \epsilon'^{\ast})\right)\right]+(e\cdot p)(q\cdot e'^{\ast})^2\left((\epsilon \cdot \epsilon'^{\ast})(e\cdot p')\right.\right.\right.\nonumber\\&&\left.\left.\left.
-(e\cdot \epsilon'^{\ast})(\epsilon\cdot p')\right)-(p\cdot \epsilon'^{\ast})(e\cdot \epsilon)(q\cdot e'^{\ast})\left((e\cdot p')(q\cdot e'^{\ast})+(e\cdot q')(e'^{\ast}\cdot p')\right)\right\}\right\}~. \label{M2}
\eea
In the following step, using Eq. \eqref{H}, together with Eq. \eqref{M2}, we will compute the time evolution of the $\rho_{ij}^{(\gamma)}$ elements.
\clearpage

\subsection{Computation of the forward scattering term }

\noindent In order to compute the forward scattering term in \eqref{Boltzmann equation2}, we need to take the expectation value of the expression
\bea
 i \left[H_{\gamma g}(t)
,\mathcal{D}^{(\gamma)}_{ij}(k)\right]&=& i \int d\mathbf{q}d\mathbf{q}'d\mathbf{p}d\mathbf{p}'(2\pi)^{3}\delta^{(3)}(\mathbf{q}'+\mathbf{p}'-\mathbf{q}-\mathbf{p})
 M_2\left[b^{\dag}_{r'}(q')b_{r}(q)a^{\dag}_{s'}(p')a_{j}(k)2p^{0}(2\pi)^{3}\delta_{is}\delta^{(3)}(\mathbf{p}-\mathbf{k})
 \right. \nonumber \\ && \left.
 -b^{\dag}_{r'}(q')b_{r}(q)a^{\dag}_{i}(k)a_{s}(p)2p'^{0}(2\pi)^{3}\delta_{js'}\delta^{(3)}(\mathbf{p}'-\mathbf{k})\right]~.\label{commut}
\eea
The expectation value of a generic operator A is defined as
\beq
 \left<A\right>=tr(\hat{\rho}^{(i)}A)\, ,
 \eeq
$\hat{\rho}^{(i)}$ being the density operator of a system of particles. Applying this definition to the case of photons, one can find the following expression for the expectation value of the product between photon creation and annihilation operators~\cite{Kosowsky:1994cy}
\beq\label{expP}
\left<a^{\dag}_{m}(\mathbf{p}')a_{n}(\mathbf{p})\right>=2p^0(2\pi)^3\delta^{(3)}(\mathbf{p}-\mathbf{p}')\rho^{(\gamma)}_{mn}(\mathbf{p}) \;,
\eeq
where $\rho^{(\gamma)}_{ij}$ is the polarization density matrix of the electromagnetic radiation
\beq \label{gamma_pol}
\rho^{(\gamma)}_{ij}=\frac{1}{2}\left(
                   \begin{array}{cc}
                     I^{(\gamma)}+Q^{(\gamma)} & U^{(\gamma)}-i V^{(\gamma)} \\
                     U^{(\gamma)}+i V^{(\gamma)} & I^{(\gamma)}-Q^{(\gamma)} \\
                   \end{array}
                 \right)~,
\eeq
$I^{(\gamma)}$, $Q^{(\gamma)}$, $U^{(\gamma)}$, and $V^{(\gamma)}$ being the
Stokes parameters. For unpolarized radiation, $Q^{(\gamma)} = U^{(\gamma)} = V^{(\gamma)} = 0$, and the parameter $I^{(\gamma)}$ describes the intensity of unpolarized radiation. For polarized radiation, $Q^{(\gamma)}$ represents the difference in the intensities of linear polarized radiation along the x and y axes, $U^{(\gamma)}$ represents the difference in the intensity of linear polarized radiation along axes rotated by $45^\circ$ with respect to the x and y axes, and $V^{(\gamma)}$ represents the difference in the intensity of the two circular polarizations.

By the same reasoning, one can find an analogous relation for gravitons
\beq\label{expGW}
\left<b^{\dag}_{m}(\mathbf{q}')b_{n}(\mathbf{q})\right>=2 q^0 (2\pi)^3\delta^{(3)}(\mathbf{q}-\mathbf{q}')\rho^{(g)}_{mn}(\mathbf{q}) \;,
\eeq
where $\rho^{(g)}_{mn}$ is the polarization density matrix of gravitons \cite{Anile,Gubitosi:2016yoq}
\beq
 \rho^{(g)}_{mn}=\frac{1}{2}\left(
                   \begin{array}{cc}
                     I^{(g)}+Q^{(g)} & U^{(g)}-i V^{(g)} \\
                     U^{(g)}+i V^{(g)} & I^{(g)}-Q^{(g)} \\
                   \end{array}
                 \right)~,
                 \eeq
in which $I^{(g)}, Q^{(g)}, U^{(g)}$ and $V^{(g)}$ are the Stokes parameters associated to gravitons; in general these are given in terms of the power-spectrum statistics of the gravitons.
As an example, let us consider $I^{(g)}$. The total tensor power-spectrum $P_h$ of gravitons is defined as

\beq
\langle h_{\mu \nu}^r(\mathbf x) h^{\mu \nu}_r(\mathbf x + \mathbf r) \rangle = \int \frac{d^3 q}{(2 \pi)^3} P_h(\mathbf q) e^{- i {\mathbf q} \cdot {\mathbf r}}\,.
\eeq
If we insert Eq. \eqref{quant} into the left-hand side of the previous equation and we use Eqs. \eqref{quant_2}  and \eqref{canonical} we find
\beq \label{I_to_P}
I^{(g)}(\mathbf q) = 2 q^0 P_h({\mathbf q}) \,.
\eeq
Now, using the expectation values \eqref{expP} and \eqref{expGW} and performing the integration over $\mathbf{p}$, $\mathbf{p}'$ and $\mathbf{q}'$, we find that the forward scattering term is given by
\bea \label{Forward_final}
i \left\langle \left[H_f(t),\mathcal{D}^{(\gamma)}_{ij}(k)\right]\right\rangle=i \, (2\pi)^{3}\delta^{(3)}(0)\int d\mathbf{q}\left(\delta_{is}\rho^{(\gamma)}_{s'j}(\mathbf{k})
-\delta_{js'}\rho^{(\gamma)}_{is}(\mathbf{k})\right)\rho^{(g)}_{rr'}(\mathbf{q})
M_2^{r,r',s,s'}(\mathbf{q}'=\mathbf{q},\mathbf{p}= \mathbf{p}'=\mathbf{k}) \;,
\eea
where the contraction between Latin indices is made with Kronecker delta. Thus, recalling the expression of $M_2$, Eq. \eqref{M2},
and inserting Eq. \eqref{Forward_final} back into Eq. \eqref{Boltzmann equation2}, after some straightforward algebra we finally get the time evolutions of the photon Stokes parameters 
in the following form 
\bea
 \dot  I^{(\gamma)}&=& 0 \,, \\
 \dot Q^{(\gamma)}&=& \frac{\kappa ^2}{k^0}\int d\mathbf{q} \,I^{(g)}(\mathbf{q})\,(q\cdot\epsilon^1 (\mathbf{k}))\,(q\cdot\epsilon^2 (\mathbf{k}))  V^{(\gamma)}~, \\
 \dot  U^{(\gamma)}&=& -\frac{\kappa^2}{2k^0}\int d\mathbf{q}\,I^{(g)}(\mathbf{q})\left[  (q\cdot\epsilon^1 (\mathbf{k}))^2-(q\cdot\epsilon^2 (\mathbf{k}))^2 \right] V^{(\gamma)} ~, \\
 \dot V^{(\gamma)}&=& \frac{\kappa ^2}{2k^0}\int{d\mathbf{q}}\,I^{(g)}(\mathbf{q})\left[ \left((q\cdot\epsilon^1 (\mathbf{k}))^2-(q\cdot\epsilon^2 (\mathbf{k}))^2\right) U^{(\gamma)}-2 \,(q\cdot\epsilon^1 (\mathbf{k}))\,(q\cdot\epsilon^2 (\mathbf{k})) Q^{(\gamma)} \right]  \;,
\eea
where dots stand for time derivatives.

Then, in order to take the integral over $d \mathbf{q}$, we fix a coordinate system where the z-axis is aligned with the three-momentum of the scattered photon $\mathbf k$ and photon polarization vectors $\bm{\epsilon}_{1}$ and $\bm{\epsilon}_{2}$ stay along x and y axes. In such a case we can write the photon and graviton kinematic variables in the following form
\bea
\mathbf{k}&=&k^0\,(0,0,1)~,\\
\mathbf{q}&=&q^0(\sin\theta'\cos\phi',\sin\theta'\sin\phi',\cos\theta')~,\\
\bm{\epsilon}_{1}&=&(1,0,0)~,\\
\bm{\epsilon}_{2}&=&(0,1, 0)~,
\eea
where $\theta'$ and $\phi'$ are the polar angles defining the direction of the three-momentum of the graviton in space. Thus, we can rewrite the previous set of equations as
\bea \label{final1}
 \dot I^{(\gamma)}&=& 0 \, ,\\
 \dot Q^{(\gamma)}&=& \frac{V^{(\gamma)}}{4k^0}\int\frac{d^{3}\mathbf{q}}{(2\pi)^3}\,q^0\sin^2\theta'\sin2\phi'\kappa^2I^{(g)}(\mathbf{q})~, \label{final2}\\
  \dot U^{(\gamma)}&=& -\frac{V^{(\gamma)}}{4k^0}\int\frac{d^{3}\mathbf{q}}{(2\pi)^3}\,q^0\sin^2\theta'\cos2\phi'\kappa^2I^{(g)}(\mathbf{q})~, \label{final3}\\
 \dot V^{(\gamma)}&=& \frac{ 1}{4k^0}\int\frac{d^{3}\mathbf{q}}{(2\pi)^3}\,q^0\sin^2\theta'[\cos2\phi'{U}^{(\gamma)}-\sin2\phi'{Q}^{(\gamma)}]\kappa^2I^{(g)}(\mathbf{q})~\;.\label{final4} 
\eea
Finally, using the results of Appendix \ref{appendixA}, we can write
\bea \label{final_1}
   \dot I^{(\gamma)}&=& 0 \, ,\\
\dot  Q^{(\gamma)}&=& \frac{V^{(\gamma)}}{2\pi k^0}\kappa^2 \bar{\rho}_{gw}\sum_{l,m}\int d^{2}\hat{\mathbf{q}}\,c^I_{lm}\sin^2\theta'\sin2\phi'Y^{m}_{l}(\theta',\phi')~, \label{final_2}\\
  \dot{{U}}^{(\gamma)}&=& -\frac{V^{(\gamma)}}{2\pi k^0} \kappa^2\bar{\rho}_{gw}\sum_{l,m}\int d^{2}\hat{\mathbf{q}}\,c^I_{lm}\sin^2\theta'\cos2\phi'Y^{m}_{l}(\theta',\phi') ~, \label{final_3}\\
\dot V^{(\gamma)}&=& \frac{ 1}{2\pi k^0} \kappa^2\bar{\rho}_{gw}\sum_{l,m}\int d^{2}\hat{\mathbf{q}}\,c^I_{lm}\sin^2\theta'[\cos2\phi' {U}^{(\gamma)}-\sin2\phi' {Q}^{(\gamma)}] Y^{m}_{l}(\theta',\phi') \;,\label{final_4}
\eea
where $\bar{\rho}_{gw}$ is the energy density of gravitons averaged over all directions, Eq. \eqref{en_dens} and $c^I_{lm}$ are the harmonic coefficients in the decomposition of 
$I^{(g)}(\mathbf{q})$ in terms of spherical harmonics, Eq. \eqref{harmonic}.

Let us briefly comment on this final set of equations: it is straightforward to verify that the source terms appearing in the right-hand sides all identically vanish when photons interact with gravitons that are characterised by a statistically isotropic power-spectrum. Thus, to achieve a nontrivial result, we need the photon to interact with an anisotropic background of gravitons. In the latter case, Q and U photon polarization states couple with the V polarization state, while the I unpolarized state remains unchanged. Notice that this result can be applied in full generality to the interactions involving gravitons and photons of whatever origin. In the next section we will give some examples applying our results to study the effect on the photon polarization due to the forward scattering with primordial gravitons generated during inflation.

\section{Forward scattering with inflationary gravitons}

\noindent Standard slow-roll models of inflation predict an isotropic power-spectrum of primordial gravitons (for a review, see, e.g., \cite{Guzzetti:2016mkm}). Therefore, in this case inflationary gravitons have no effect in Eqs. \eqref{final_2}, \eqref{final_3}, and \eqref{final_4}. In this section, we will briefly review some alternative models of inflation where a certain level of anisotropy in the tensor power-spectrum is generated and, using Eq. \eqref{I_to_P}, we will link the power-spectrum predicted by these models to the results found at the end of Sec. II. At the end of the section we will provide a general estimate of the effects on CMB polarization.

\subsection{Anisotropic solid inflation}

\noindent Anisotropic solid inflation is a novel inflationary model studied in Refs.~\cite{Peloso:2013, Bartolo:2014xfa,Akhshik:2014}, based on the original model of solid inflation~\cite{Endlich:2013jia}. According to this model, the inflationary period is driven by a configuration which behaves like a solid: the space is fragmented into cells whose location is defined by a triplet of scalar fields $\phi^I(t, \mathbf{x})$, where $I= 1$, 2 or 3. The three scalars can be viewed as the three coordinates that give the position, at time $t$, of the cell element that, at the time $t = 0$, was in the position 
$\mathbf{x}$. At the background level one has

\begin{equation}
\langle \phi^I \rangle = x^I \quad , \quad I = 1, 2, 3.
\end{equation}
From the previous equation we understand that the scalar fields $\phi^I$ are time-independent at the background level and give a sort of  \textquotedblleft average position\textquotedblright of each cell. In order to require isotropy and homogeneity of the background, in the Lagrangian of the theory the following internal symmetries are imposed

\begin{equation}
\label{cond1}
\phi^{I} \rightarrow \phi^I + C^I
\end{equation}
and
\begin{equation}
\label{cond2}
\phi^I \rightarrow O^I_J \phi^J \quad , \quad O^I_J \in SO(3) \,.
\end{equation}
The most general action consistent with the previous symmetries and minimally coupled to gravity is given by \cite{Endlich:2013jia}
\begin{equation}
\label{solid-action}
S = \int d^4 x \sqrt{-g} \left\{ \frac{M_P^2}{2} R + F[X, Y, Z] \;,
\right\} \, ,
\end{equation}
where
\begin{equation}
X \equiv {\rm Tr } \,  B =  B^{ii} \;\;\;,\;\;\; Y \equiv \frac{{\rm Tr } \,  \left( B^2 \right)}{\left( {\rm Tr  } \, B \right)^2} \;\;\;,\;\;\;
 Z \equiv \frac{{\rm Tr } \left( B^3 \right)}{\left( {\rm Tr  } \, B  \right)^3} \;,\quad  B^{IJ} \equiv g^{\mu \nu} \partial_\mu \phi^I \partial_\nu \phi^J \, .
\end{equation}
Writing down the background cosmological equations, the slow-roll parameters turn out to be \cite{Endlich:2013jia,Bartolo:2014xfa}
\begin{equation}
\epsilon = \frac{X F_X}{F} \quad , \quad
\eta = 2\left( \epsilon -1 - \frac{X^2 F_{XX}}{X F_X} \right) \, ,
\end{equation}
where $F_X = \partial F/ \partial X$ and the same for Y and Z.

The scalar field perturbations are given by
\begin{equation}
\phi^I = x^I + \pi^I(t, \bf x) \,.
\end{equation}
In particular, it is possible to decompose the perturbations $\pi^I(t, \bf x)$ into a transverse and a longitudinal part, as
\begin{equation}
\label{pi-decompose}
\pi^I(t, \mathbf x) = {\partial_I} \pi_L(t, \mathbf x) + \pi^I_T(t, \mathbf x)\, ,\qquad \partial_I\pi^I_T = 0 \,.
\end{equation}
The field $\pi_L(t, \mathbf x)$ labels the ``phonons'' for the longitudinal fluctuations of the solid. The sound speeds of longitudinal and transverse excitations are given 
by \cite{Endlich:2013jia,Bartolo:2014xfa}
\begin{equation}
c_L^2  \equiv  1+ \frac{2F_{XX} X^2}{3F_X X} + \frac{8(F_Y + F_Z)}{9 F_X X} \, ,
\end{equation}
and
\begin{equation}
c_T^2 = 1 + \frac{2(F_Y + F_Z)}{3 X F_X} =
\frac{3}{4} \left(1+ c_L^2 - \frac{2\epsilon}{3} + \frac{\eta}{3}\right) \, .
\end{equation}
The anisotropic version of this model is achieved introducing a preferred direction in the background metric; for example, we can consider a Bianchi type-I background geometry with residual 2d isotropy
\begin{eqnarray}
&& d s^2 =  - d t^2 + a^2 \left( t \right) d x^2 + b^2 \left( t \right) \left[ d y^2 + d z^2 \right]    \,\,, \nonumber\\
&& a \equiv {\rm e}^{\alpha - 2 \sigma} \;\;,\;\;
b \equiv {\rm e}^{\alpha +  \sigma}  \,\,, 
\label{bianchi}
\end{eqnarray}
where the field $\sigma$ labels the anisotropy. Here the x-axis is labeled as the preferred direction. Einstein's equations for this kind of background applied to solid inflation are given by \cite{Bartolo:2014xfa}

\begin{align}
H^2 -\dot \sigma^2  =& - \frac{F}{3 M_P^2} \, , \\
\dot H + 3 \dot \sigma^2  =&  \frac{e^{4 \sigma} + 2 e^{-2 \sigma}}{3 M_P^2} e^{-2\alpha} F_X \, , \\
\ddot \sigma + 3 H \dot \sigma  =& \frac{2 ( e^{4 \sigma} -  e^{-2 \sigma} )}{3 M_P^2}
e^{-2\alpha} F_X  - \frac{4 e^{6\sigma} ( 1-e^{6 \sigma}  ) F_Y}{(2+ e^{6 \sigma})^3 M_P^2}
- \frac{6 e^{6\sigma} (1-  e^{12 \sigma})  F_Z }{( 2+ e^{6 \sigma}  )^4 M_P^2 }\label{last} \, .
\end{align}
If we rewrite Eq. \eqref{last} in the small anisotropy limit (i.e. $\sigma \ll 1$), it becomes
\begin{equation}\label{sigma_eq}
\ddot \sigma + 3 H \dot \sigma + 4 \epsilon H^2 c_T^2 \sigma = 0 \,.
\end{equation}
Assuming nearly constant values of $\epsilon$ and $c_T$ one can solve Eq. \eqref{sigma_eq} finding~\cite{Bartolo:2014xfa}
\begin{equation}
\label{sigma-t}
\sigma(t) \simeq \sigma_1 e^{-\int dt\, \left[ (3- (2+c_L^2) \epsilon ) \right] H }
+ \sigma_2  e^{-\int dt\, \frac{4}{3} c_T^2 \epsilon H  }\, ,
\end{equation}
where $\sigma_1$ and $\sigma_2$ are two constants.

From Eq. \eqref{sigma-t} we understand that the general solution is a superposition of two kinds of solutions: the solution proportional to $\sigma_1$ is fast-decaying, while the solution proportional to 
$\sigma_2$ is slowly decaying. The result is that, immediately after the beginning of inflation, only the second solution survives. Moreover, if inflation does not last longer than a time $1/\sqrt \epsilon c_T$, a residual anisotropic deformation of the background is preserved~\cite{Bartolo:2014xfa}; thus, solid inflation is not efficient in diluting the initial anisotropy, contrary to what happens in the standard slow-roll inflationary scenario
\footnote{In fact, in slow-roll models of inflation with a Bianchi type-I background Eq. \eqref{sigma_eq} simplifies, becoming
$$
\ddot \sigma + 3 H \dot \sigma  = 0 \,.
$$
The solution of this equation reads
$$
\sigma(t) \simeq \sigma_0 e^{-\int dt\, 3H} \, .
$$
Thus, a rapid dilution of all the initial small anisotropy follows. This result is also known as \textit{cosmic no hair theorem}, and states that slow-roll inflation rapidly erases 
all kinds of anisotropies. For this reason, a Bianchi type-I slow-roll model of inflation is not sufficient to lead to anisotropies in the primordial perturbations.
}.  This is due to the fact that to produce inflation, the solid must be insensitive to the spatial expansion, but, at the same time, this makes it rather inefficient in erasing anisotropic deformations of the geometry. In fact,  Ref.~\cite{Bartolo:2014xfa} shows that it is rather general to expect anisotropic evolution in these scenarios. For this reason, during a solid inflationary period, signatures of primordial anisotropies can be left imprinted into primordial perturbations~\cite{Bartolo:2014xfa,Endlich:2013jia}. In particular, in Ref. \cite{Akhshik:2014} the anisotropy in the power-spectrum statistics of primordial gravitons has been computed. The final result is
\begin{equation}
\label{Ph-total}
P^{solid}_h(\mathbf q) = P_h^{(0)}(q) \left[1 + \frac{16 N \sigma F_Y}{9 \epsilon c_L^5 F}  \left[ \epsilon c_L^5 (1 - 3 \cos^2 \alpha)
 {+} 2  \sigma N \frac{F_Y}{F}  \sin ^{{4}} \alpha \right] \right] \;,
\end{equation}
where $P^{(0)}_h(q) = 4 H^2/ q^3 M^2_{Pl}$ is the total power-spectrum of gravitons in the standard slow-roll inflationary models, $N=-\ln (-q \eta_e)$ is the number of e-folds when the mode ${\bf q}$ leaves the horizon until the end of inflation, $\alpha$ is the angle between the preferred direction and the momentum $\mathbf q$ of the graviton.

Now, let us assume that the preferred direction is completely general in the (x,y) plane \footnote{Notice that, in principle, we could choose completely general $\hat n$. However, from angular integrals in Eqs. \eqref{final2}, \eqref{final3} and \eqref{final4} it follows that it is sufficient to introduce in $I^{(g)}$ a dependence on the polar angle $\phi'$ to achieve a non-trivial result.}, so that 
$\hat n=(\cos \phi, \sin \phi , 0)$. The angle $\alpha$ between the preferred direction $\hat n$ and the generic graviton momentum $\hat q =(\sin\theta'\cos\phi',\sin\theta'\sin\phi',\cos\theta')$ is given by
\begin{equation}
\cos \alpha = \hat n \cdot \hat q = \cos \phi' \sin \theta' \cos \phi  + \sin \phi' \sin \theta' \sin \phi \, ,
\end{equation}
\begin{equation}
\sin^2 \alpha = 1- \cos^2 \alpha = 1 -  \cos^2 \phi' \sin^2 \theta' \cos^2 \phi  - \sin^2 \phi' \sin^2 \theta' \sin^2 \phi - 2 \cos \phi' \sin^2 \theta' \cos \phi \sin \phi' \sin \phi \, .
\end{equation}
Thus, apart from some coefficients, the integrals we have to compute turn out to be
\begin{equation}\label{I1}
I_1 = \frac{1}{4}\int\frac{d^{3}\mathbf{q}}{(2\pi)^3}\,q^0\sin^2\theta'\sin2\phi'\kappa^2I_{solid}^{(g)}(\mathbf{q})~ = - B \,  \sin(2\phi) \,,
\end{equation}
and
\begin{equation}\label{I2}
I_2 = \frac{1}{4} \int\frac{d^{3}\mathbf{q}}{(2\pi)^3}\,q^0\sin^2\theta'\cos2\phi'\kappa^2 I_{solid}^{(g)}(\mathbf{q})~ = - B \, \cos(2\phi) \,,
\end{equation}
where
\begin{equation}
B =  \kappa^2 \bar \rho_{gw} \frac{4 N \sigma F_Y}{9 \epsilon c_L^5 F} \left[\frac{8}{5} \epsilon c_L^5
{+} \frac{128}{105}  \sigma N \frac{F_Y}{F} \right] \,,
\end{equation}
$\bar \rho_{gw}$ being the energy density of primordial gravitons averaged over directions, as defined in Eq. \eqref{en_dens}.

At the end, inserting Eqs. \eqref{I1} and \eqref{I2} into Eqs. \eqref{final2}, \eqref{final3} and \eqref{final4} we obtain
\bea
\dot Q^{(\gamma)}&=&\frac{I_1}{k^0} V^{(\gamma)}~, \label{Q_solid}\\
\dot {U}^{(\gamma)}&=& -\frac{I_2}{k^0} V^{(\gamma)}~, \label{U_solid}\\
\dot V^{(\gamma)}&=& \frac{1}{k^0} \left[I_2{U}^{(\gamma)}- I_1 {Q}^{(\gamma)}\right]~ \label{V_solid}.
\eea

\subsection{Squeezed non-Gaussianity anisotropic imprint}

\noindent Squeezed tensor bispectra can lead to anisotropic modulations of the tensor power-spectrum in some inflationary scenarios, similarly to the effect induced in the curvature power-spectrum by the so-called tensor ``fossils''  (see~\cite{Masui:2010, Jeong:2012,Dai:2013, Jeong:2013, Brahma:2014,Bordin:2016, Ricciardone:2017}). This can happen, e.g., in models of inflation where space-time diffeomorphisms are spontaneously broken (see, e.g.,~\cite{Bordin:2016, Ricciardone:2017}). Here spontaneous breaking means that one or more scalar fields driving inflation admit a background value which is not invariant under a generic space-time reparametrization.

In general, the squeezed limit of the three-gravitons bispectrum is given by (see, e.g., \cite{Ricciardone:2017})
\begin{equation} \label{squeezed}
\langle h_{s_1}(\mathbf{q_1}) h_{s_2}(\mathbf{q_2}) h_{s}(\mathbf{Q}) \rangle \xrightarrow[q_1 \simeq q_2]{Q \rightarrow 0} (2 \pi)^3 \delta^{(3)}(\mathbf{q_1 + q_2+Q})  P_h(Q) P_h(q) \left(\frac{3}{2}+ f_{NL}\right) \epsilon^s_{ij}(\mathbf{Q}) \, \mathbf{\hat q_1^i \hat q_{2}^j} \, \delta_{s_1 s_2} \;,
\end{equation}
where $\mathbf Q$ and $\mathbf q$ are the momenta of the long and short modes, respectively. The parameter $f_{\rm NL}$ characterizes how much we are violating the so-called Maldacena's consistency relation (see Ref. \cite{Maldacena:2003}), due to the spontaneously breaking of space-time reparametrizations.

It is possible to show that a single soft graviton $ \tilde h_s({\mathbf Q})$ is able to modulate the tensor power-spectrum as follows~\cite{Ricciardone:2017}
\beq
P_h(\mathbf q)|_h= P_h(q)^{(0)} + \tilde h_s({\mathbf Q}) \frac{\langle h_{s_1}(\mathbf{q_1}) h_{s_2}(\mathbf{q_2}) h_{s}(\mathbf{Q}) \rangle'}{P_h(Q)}\, ,
\eeq
where $P_h^{(0)}$ is the unmodulated total tensor power-spectrum and the prime $'$ means that we have to drop the Dirac-delta in the corresponding expression. 
Moreover, if we want to look for this modulation at a given position $\mathbf x$ in a cube of volume $V$ of physical space, we should consider the cumulative effect of all soft graviton modes with minimum wavelength $\lambda_L = {V}^{1/3}$. This leads to a local quadrupolar anisotropy in the total tensor power-spectrum which can be parametrized as
\beq\label{modulation}
P_h^{squeezed}(\mathbf q, \mathbf x) = P_h^{(0)}(q) \left( 1 + \gamma_{ij}(\mathbf{x}) \hat q^i \hat q^j \right)\, ,
\eeq
where
\beq
\gamma_{ij}(\mathbf{x}) = \frac{f_{NL}}{V_L}\int_{|Q|<Q_L} d^3  Q \,\, e^{i{\mathbf{Q \cdot x}}} \,\, h_s(\mathbf Q) \epsilon^s_{ij}(\mathbf Q)\, .
\eeq
From its definition $\gamma_{ij}(\mathbf{x})$ depends on the position in space and is a stochastic Gaussian tensor field, with variance given by
\beq\label{var}
\langle \gamma_{ij}(\mathbf{x})\gamma_{ij}(\mathbf{x}) \rangle = \frac{f_{NL}^2}{V_L^2}  \int_{|Q|<Q_L} d^3Q \, P_h(\mathbf Q) \,.
\eeq
From Eq. \eqref{modulation} we understand that $\gamma_{ij}(\mathbf{x})$ labels the local preferred directions of the tensor power-spectrum, due to the modulation provided by long modes. For simplicity, we can assume that $\gamma_{ij}(\mathbf{x})$ is a constant, thus $\gamma_{ij}(\mathbf{x}) = \gamma_{ij}$. In this case $\gamma_{ij}$ is a three-dimensional symmetric constant matrix whose entries are proportional to $f_{\rm NL}$ and fix the form of the quadrupolar angular dependence. Notice that, a priori, $\gamma_{ij}$ takes a random value extracted from a Gaussian distribution with variance given by Eq. \eqref{var}. So, the precise angular dependence of the quadrupolar anisotropy is not completely fixed by the theory, but depend on which particular realization 
$\gamma_{ij}$ is fixed during inflation.

Now, we take the modulated power-spectrum
\beq\label{modulation2}
P_h^{squeezed}(\mathbf q) = P_h^{(0)}(q) \left( 1 + \bar \gamma_{ij} \hat q^i \hat q^j \right)\, ,
\eeq
with a fixed value  $\bar \gamma_{ij}$, and we substitute Eq. \eqref{modulation2} into Eqs. \eqref{final2}, \eqref{final3}, and \eqref{final4}.
Recalling Eq. \eqref{I_to_P} and performing some simple integrals, we finally get
\bea
 \dot Q^{(\gamma)}&=& \bar \gamma_{12} \frac{4}{15}  \frac{ \kappa^2 \bar \rho_{gw}}{k^0} V^{(\gamma)}\, \,, \label{Q_squ}\\
\dot {U}^{(\gamma)}&=& - \left(\bar \gamma_{11} - \bar \gamma_{22} \right) \frac{2}{15}  \frac{\kappa^2 \bar \rho_{gw}}{k^0} V^{(\gamma)} ~, \label{U_squ}\\
 \dot V^{(\gamma)}&=& \frac{2}{15}  \frac{\kappa^2 \bar \rho_{gw}}{k^0} \left[ \left(\bar \gamma_{11} - \bar \gamma_{22} \right)U^{(\gamma)} - 2  \bar \gamma_{12} Q^{(\gamma)} \right]~. \label{V_squ}
\eea

\subsection{Effects on CMB polarization}

\noindent It is of great cosmological interest to evaluate the effect of the (quantum) Boltzmann equations we derived on the CMB photon polarization; in fact, the CMB represents one of the most important and studied cosmological sources of photons in the universe (see e.g. Ref \cite{Planck}).

Before the recombination epoch CMB photons are basically unpolarized, thus $Q = U = V = 0$ (see, e.g., \cite{dodelson:2003}). In this case, it is straightforward to show that no new physics is provided by the Boltzmann equations we derived. Then, at the time of recombination, a small amount of $Q$ polarization modes due to the Compton scattering of photons with baryons is formed 
\cite{dodelson:2003}. Thus, after the recombination epoch, we start with initial conditions where only the $Q$ polarization mode is nonzero, i.e. $Q(0) = Q_{init}$ and $U(0)= V(0) = 0$. In such a case, according to our Boltzmann equations, $V$ modes are initially coupled only with $Q$ modes, through a set of differential equations like
\bea \label{Eq1}
\dot V &= & \left(\frac{\kappa^2 \bar \rho_{gw} }{k^0} A \right) Q\, ,\\
\dot Q &= & - \left(\frac{\kappa^2 \bar \rho_{gw} }{k^0} A \right) V\, ,
\label{Eq2}
\eea
where 
\beq
A =  \frac{1}{2\pi } \sum_{l,m}\int d^{2}\hat{\mathbf{q}}\,c^I_{lm}\sin^2\theta'\sin2\phi'Y^{m}_{l}(\theta',\phi')
\eeq
is a dimensionless parameter depending on the underlying theory. In Eqs. \eqref{Eq1} and \eqref{Eq2} we have dropped the $U$ modes. In fact, in our Boltzmann equation $U$ modes vanish at the beginning of the time evolution and they are coupled only with $V$ modes, hence they cannot be produced until a reasonable amount of $V$ modes is produced in turn. If we neglect the time dependence of $k^0$ and $\bar \rho_{gw}$ due to the CMB gravitational redshift, this kind of coupled set of differential equations can be easily solved leading to the oscillatory behavior

\bea
V = Q_{init} \sin(\omega t)\, \label{sol1},\\
Q = Q_{init} \cos(\omega t)\, \label{sol2},
\eea
where
\beq \label{omega}
\omega = \frac{\kappa^2 \bar \rho_{gw} }{k^0}  A \, \\
\eeq
is the frequency of the oscillation.

The result is that the value of $Q$ starts to decrease like a cosine, while the value of $V$ grows like a sine. At a certain time, when the value of $V$ becomes important, the coupling between $V$ and $U$ modes should be taken into consideration and can potentially lead also to the generation of $U$ modes, modifying our general solution. 
However, let us neglect for simplicity the $U$ modes in all the discussion. This is equivalent to make a fine-tuning in the parameters of the models we considered, in order to decouple $U$ and $V$ modes in the Boltzmann equations \footnote{In the case of \textit{anisotropic solid inflation} it is enough to choose $\phi = \pi/4$ in Eqs. \eqref{I1} and \eqref{I2}; in the case of \textit{squeezed non-Gaussianity theories} we need  $\left(\bar \gamma_{11} - \bar \gamma_{22} \right) = 0$ in Eqs. \eqref{U_squ} and \eqref{V_squ}.}.

From Eqs. \eqref{Eq1} and \eqref{Eq2} we understand that the effect on CMB polarization is greater for photons with smaller wave number $k^0$, thus we consider CMB photons with comoving frequency of 1 GHz ($k^0 = 2 \pi  f $), which is the order of magnitude of the smallest CMB frequency that has been measured~\footnote{More precisely, the lowest measured CMB frequency corresponds to 0.6 GHz as measured by the TRIS instrument (see e.g. Ref. \cite{TRIS:2008}).}. From the constraint on the total energy density of gravitational waves today, provided by 
nucleosynthesis ($\bar \rho_{gw} \lesssim 10^{-5} \rho^o_{crit}$, where $\rho^o_{crit}$ is the critical energy density of the universe today, see Ref. \cite{Allen}) and from the definition of $\kappa^2 = 16 \pi G$ it follows that, at the time of recombination epoch (at redshift $z \approx 1100$), we have
\beq
\omega_{rec}(f = 1 \, \mbox{GHz}) \lesssim 5 \times 10^{-41}  A  \,\,s^{-1} \, .
\eeq
From this constraint it follows that, if we take $A \sim 1$, then, one year immediately after the recombination epoch, we would obtain
\beq
V \lesssim 10^{-34} Q_{init} \, .
\eeq
If we neglect gravitational red-shift and we integrate the effect from the recombination epoch until today ($\sim 10^9$ yr) we have
\beq \label{V_strength}
V \lesssim 10^{-25} Q_{init} \, .
\eeq
In this last case, neglecting in $10^9$ yr the effect of gravitational redshift on the frequency $k^0$ and on the energy density $\bar \rho_{gw}$ is indeed not a good approximation. However, if we account for it, we expect that the upper bound on $V$ is even smaller than the one shown in Eq. \eqref{V_strength} \footnote{In fact, the energy density of gravitons and the physical wave number of CMB photons in terms of gravitational redshift $z$ are given respectively by
$$\rho_{crit}(z) = (1+z)^4 \rho^o_{crit}$$
and
$$k^0(z) = (1+z)\, k^0_{com} \, .$$
Thus, the parameter $\omega(z)$ in Eq. \eqref{omega} scales as
$$\omega(z) = (1+z)^3 \omega_o \,.$$
Inserting the latter equation into Eqs. \eqref{Eq1} and \eqref{Eq2} we understand that the strength of the coupling between $Q$ and $V$ modes in the past is larger than the one today. Thus, if we neglect the effect of gravitational redshift, we would overestimate the final amount of $V$ modes today.
}.
Therefore, these estimations show that the forward scattering between CMB photons and (primordial) anisotropic gravitons leads to the production of circular polarization in the CMB, but this in general is very inefficient. This brief analysis suggests that unfortunately the CMB does not seem to be the best source of photons to be used in searching for a signature of anisotropic (primordial) gravitons, and one should look for other kinds of sources.

\section{Discussion and Conclusions}

\noindent In this paper we used a QFT approach to study the forward scattering of photons with gravitons, focusing on the effect that this scattering has on photon polarization. 
We derived fully general (quantum) Boltzmann equations which display a coupling among the Q, U linear polarization states and the V circular polarization state of photons. In fact, these couplings are not vanishing only if photons interact with gravitons that have anisotropies in their power-spectrum statistics. As an application of our general results we have considered some models of inflation where primordial anisotropic gravitons are generated and we linked our Boltzmann equations to these models. Finally, we evaluated the effect of a primordial anisotropic background of gravitons on the CMB polarization. We saw that in general the effect on the CMB is expected to be very small in a way that it is probably impossible to measure it via CMB polarization measurements. However, we have to be open-minded and think about alternative scenarios where  \textquotedblleft artificial\textquotedblright polarized photons can be employed: in this case, playing with the initial values of the Stokes parameters and with the frequency of the photons we could enhance the effects on the photons polarization. Then, we could use controlled photons to probe statistical anisotropies in (primordial) backgrounds of gravitational waves, e.g. constraining free parameters of alternative models of inflation. More in general, we could use our result to search for any source of anisotropic gravitons in the universe. In the future we could use this new tool to get insight into the physics of gravitational waves and provide an innovative way to look for gravitational-wave events.

\section*{Acknowledgments}

\noindent M. Zarei would like to thank INFN and department of Physics and Astronomy \textquotedblleft G. Galilei\textquotedblright  at University of Padova for financial support. M. Zarei  would also like to thank all the staff for their warm hospitality during his visit in Padova. Moreover, we would like to thank Massimo Pietroni, Angelo Ricciardone,  Pierpaolo Mastrolia and Roohollah Mohammadi for useful discussions. We also acknowledge partial financial support by ASI Grant No. 2016-24-H.0. Ê


\begin{appendices}
\section{: The energy density of gravitational waves} \label{appendixA}

\noindent In this Appendix we derive the energy density of gravitons (i.e. gravitational waves) in terms of gravitational Stokes parameter $I^{(g)}$. Starting from the Lagrangian  \eqref{Lg0} we can find the energy density of gravitational waves as \footnote{Normally it is impossible to define a full nonlinear local energy-momentum tensor for the gravitational field $g_{\mu \nu}$. The definition \eqref{energy} can be used in our context because we work in the weak field approximation, as stated in Eq. \eqref{weak_field} (see Ref. \cite{maggiore2008gravitational} for more details).}
\bea \label{energy}
\rho_{gw}=\frac{1}{2} \braket{\dot{h}_{\mu \nu}  \,\, \dot{h}^{\mu \nu}}~.
\eea
By inserting
\bea
\dot{h}_{\mu\nu}(x)=\frac{i}{2}\int\frac{d^{3}q}{(2\pi)^{3}}\sum_{r=+,\times}\left[b^{(r)}_{\mathbf{q}}\,h^{(r)}_{\mu\nu}\,e^{iqx}-
b^{(r)\,\dag}_{\mathbf{q}}\,h^{(r)\,\ast}_{\mu\nu}\,e^{-iqx}\right]~,
\eea
we have
\bea
\rho_{gw}=-\frac{1}{8} \left<\int\frac{d^{3}q}{(2\pi)^{3}} \int\frac{d^{3}q'}{(2\pi)^{3}}\sum_{r,r'}\left[b^{(r)}_{\mathbf{q}}\,h^{(r)}_{\mu\nu}\,e^{iqx}-
b^{(r)\,\dag}_{\mathbf{q}}\,h^{(r)\,\ast}_{\mu\nu}\,e^{-iqx}\right] \left[b^{(r')}_{\mathbf{q'}}\,h^{\mu\nu(r')}\,e^{iq'x}-
b^{(r')\,\dag}_{\mathbf{q'}}\,h^{ \mu\nu(r')\ast}\,e^{-iq'x}\right]\right> ~,
\eea
which is further simplified as
\bea
\rho_{gw}&=&-\frac{1}{8} \int\frac{d^{3}q}{(2\pi)^{3}} \int\frac{d^{3}q'}{(2\pi)^{3}}\sum_{r=+,\times}\sum_{r'=+,\times}\left\{h^{(r)}_{\mu\nu}\,h^{\mu\nu(r')}\,e^{i(q+q')(x)}\,\braket{b^{(r)}_{\mathbf{q}}\,b^{(r')}_{\mathbf{q'}}} - h^{(r)}_{\mu\nu}\,h^{\mu\nu (r') \ast }\,e^{i(q-q')(x)}\,\braket{b^{(r)}_{\mathbf{q}}\,b^{(r')\,\dag}_{\mathbf{q'}}} \right.\nonumber\\
&&\left. -h^{(r)\,\ast}_{\mu\nu}\,h^{\mu\nu(r')}\,e^{i(q'-q)(x)}\,\braket{b^{(r) \dag}_{\mathbf{q}}\,b^{(r')\,}_{\mathbf{q'}}}+ h^{(r)\,\ast}_{\mu\nu}\,h^{\mu\nu(r') \ast }\,e^{-i(q'+q)(x)}\,\braket{b^{(r) \dag}_{\mathbf{q}}\,b^{(r')\,\dag}_{\mathbf{q'}}}\right\} \nonumber\\
 &=&\frac{1}{8} \int\frac{d^{3}q}{(2\pi)^{3}} \int\frac{d^{3}q'}{(2\pi)^{3}}\sum_{r=+,\times}\sum_{r'=+,\times} h^{(r)\,\ast}_{\mu\nu}\,h^{\mu\nu(r')}\,e^{i(q'-q)(x)}\,\braket{b^{(r) \dag}_{\mathbf{q}}\,b^{(r')\,}_{\mathbf{q'}}} \nonumber\\
&=&\frac{1}{8} \int\frac{d^{3}q}{(2\pi)^{3}} \int\frac{d^{3}q'}{(2\pi)^{3}}\sum_{r=+,\times}\sum_{r'=+,\times}  h^{(r)\,\ast}_{\mu\nu}\,h^{\mu\nu (r')}\,e^{i(q'-q)(x)}\, \left(2q'^0 (2\pi)^3 \delta^{(3)}(\mathbf{q'}-\mathbf{q}) \rho_{r r'}(\mathbf{q'})\right)~.
\eea
Now, using Eqs. \eqref{expGW} and  \eqref{canonical}, we get
\bea
\rho_{gw}&=&\frac{1}{4} \int\frac{d^{3}q}{(2\pi)^{3}} \sum_{r=+,\times}\sum_{r'=+,\times}  q^0 \rho_{r r'}(\mathbf{q}) \delta_{r,r'}= \frac{1}{4} \int\frac{d^{3}q}{(2\pi)^{3}}\,  q^0 \left[ \rho_{+ +}(\mathbf{q})+\rho_{\times \times}(\mathbf{q})\right] \nonumber\\
&= &\frac{1}{4} \int\frac{d^{3}q}{(2\pi)^{3}}\,q^0 I^{(g)}(\mathbf{q})~.
\eea
We can expand $I^{(g)}(\mathbf{q})$ in terms of spherical harmonics as \cite{Kato:2015bye}
\beq \label{harmonic}
I^{(g)}(\mathbf{q})=I^{(g)}(q^0)\sum_{l,m}c^I_{lm}Y^{m}_{l}(\theta',\phi')~.
\eeq
Therefore, one can write
\bea
\rho_{gw}&=&\frac{\pi}{2} \int df f^3 I^{(g)}(f)\sum_{l,m}\int d^{2}\hat{\mathbf{q}}\,c^I_{lm}Y^{m}_{l}(\theta',\phi')~,
\eea
where $q^0=2\pi f$. The isotropic part of gravitational waves energy density, $\bar{\rho}_{gw}$, is given by
\beq
\bar{\rho}_{gw}=\frac{\pi}{2}\frac{c^I_{00}}{\sqrt{4\pi}} \int df f^3 I^{(g)}(f)\int d^{2}\hat{\mathbf{q}}~,
\eeq
and, after normalizing the monopole moment as $c^I_{00}=\sqrt{4\pi}$ and performing the angular integral, we finally get
\beq  \label{en_dens}
\bar{\rho}_{gw}= 2 \pi^2 \int df f^3 I^{(g)}(f)~.
\eeq

\end{appendices}


\begingroup 
  \makeatletter
  \let\ps@plain\ps@empty
  \makeatother
  \bibliography{bibl}
\endgroup


\end{document}